\documentclass[fleqn,usenatbib]{mnras}

\usepackage{newtxtext,newtxmath}

\usepackage[T1]{fontenc}
\usepackage{ae,aecompl}

\usepackage{graphicx}	
\usepackage{amsmath}	
\usepackage{amssymb}	
\usepackage{enumitem}
\usepackage{etoolbox}
\usepackage{siunitx}
\usepackage{listings}
\usepackage{graphicx}
\usepackage{float}
\usepackage{subfigure}
\graphicspath{ {images/} }
\usepackage{physics}
\usepackage{courier}

\newcommand{\mc}{\mathcal}
\newcommand{\be}{\begin{equation}}
\newcommand{\ee}{\end{equation}}
\newcommand{\bea}{\begin{eqnarray}}
\newcommand{\eea}{\end{eqnarray}}
\newcommand{\nn}{\nonumber}
\newcommand{\ti}{\times}

\title[]{Constraints on Massive Axion-Like Particles from X-ray Observations of NGC1275}

\author[]{
Linhan Chen$^1$, Joseph P. Conlon$^2$
\\
$^{1}$ Princeton University, 4108 Frist Center, Princeton, NJ 08544, USA \\
$^{2}$ Rudolf Peierls Centre for Theoretical Physics, 1 Keble Road, Oxford, OX1 3NP, UK
}

\pubyear{2017}

\begin{document}
\label{firstpage}
\pagerange{\pageref{firstpage}--\pageref{lastpage}}
\maketitle

\begin{abstract}
If axion-like particles (ALPs) exist, photons can convert to ALPs on passage through regions containing magnetic fields. The magnetised intracluster medium of large galaxy clusters provides a region that is highly efficient at ALP-photon conversion. X-ray observations of Active Galactic Nuclei (AGNs) located within galaxy clusters can be used to search for and constrain ALPs, as photon-ALP conversion would lead to energy-dependent quasi-sinusoidal modulations in the X-ray spectrum of an AGN.
We use \textit{Chandra} observations of the central AGN of the Perseus Cluster, NGC1275, to place bounds on massive ALPs up to $m_a \sim 10^{-11} {\rm eV}$, extending previous work that used this dataset to constrain massless ALPs.
\end{abstract}




\section{Introduction}
Axions are a hypothetical extension of the Standard Model, originally motivated by providing an appealing solution to the strong CP problem of QCD (\cite{PecceiQuinn, Weinberg, Wilczek}). A recent review of axion physics is \cite{Marsh}. 
While the original QCD axion requires a coupling to the strong force, it is also interesting to consider more general \emph{axion-like particles} (ALPs) that couple only to electromagnetism. Such ALPs arise generally in string compactifications (for example, see \cite{Conlon:2006tq, Svrcek:2006yi, Cicoli:2012sz}). An ALP $a$ interacts with photons via the Lagrangain coupling:
\bea
a g_{a\gamma\gamma}\vec{E}\cdot\vec{B} & \in & \mc{L},
\eea
where $g_{a\gamma\gamma}$ is a constant that parametrizes the strength of the coupling and $\vec{E},\vec{B}$ are the electric and magnetic fields, respectively. While we refer in this paper to ALPs, the physics is also relevant for photophilic models of the QCD axion in which the mass is much smaller (or the photon coupling significantly enhanced) compared to naive expectations, such as (\cite{Farina, Agrawal, Agrawal2}).

As they attain masses only by non-perturbative effects, ALPs  naturally have extremely small masses. The relevant physics is then described by the Lagrangian
\bea
\mathcal{L} = \frac{1}{2} \partial_{\mu} a \partial^{\mu} a + \frac{1}{2} m_a^2 a^2 + a g_{a\gamma\gamma}\vec{E}\cdot\vec{B}.
\eea\par 
The ALP-photon coupling produced by the $a\vec{E}\cdot\vec{B}$ interaction implies that, within a background magnetic field, the ALP state $a$ has a 2-particle interaction with the photon $\gamma$. Under this mixing, the `mass' eigenstate of the Hamiltonian is a mixture of the photon and ALP `flavour' eigenstates, causing oscillation between the modes in a way analogous to neutrino oscillations. The dynamics of the mixing are discussed in \cite{Sikivie:1983ip} and \cite{RS}, and we briefly review it in section 2.

ALP-photon conversion is enhanced by large magnetic field coherence lengths. As it extends over megaparsec sccales and contains coherence lengths up to tens of kiloparsecs, 
this makes the intracluster medium of
galaxy clusters particularly efficient ALP-photon converters (\cite{Burrage:2009mj, Angus:2013sua, Conlon:2013txa, Powell:2014mda, Day2015, Schlederer:2015jwa, Conlon:2015uwa, Jennings2017}). For X-ray point sources that are located in or behind a cluster, this conversion can produce quasi-sinuosoidal modulations in the spectrum of the source, that can be used to constrain ALP parameter space (\cite{Wouters:2013hua, ConGroup, ConGroup2, Marsh:2017yvc, Conlon:2017ofb}) (see \cite{WoutersGammas2012, Ajello, Payez, Montanino2017} for some related work in different wavebands).

Generally, the cluster magnetic field is expected to be turbulent and multi-scale, with a characteristic range of coherence lengths in the 1- 10kpc range with a central magnetic field strength in the 2- 30 $\mu$G range (larger for cool core clusters). The magnetic field properties are measured using Faraday Rotation Measures (see e.g. \cite{Bonafede:2015pka}).
The radial dependence of the overall magnitude of the cluster magnetic field $\left\lvert\vec{B}(r)\right\rvert$ is conventionally related to the electron density $n_e(r)$ as
\bea\label{B}
\left\lvert\vec{B}(r)\right\rvert=\left\lvert\vec{B}_0\right\rvert\left(\frac{n_e(r)}{n_e(0)}\right)^{\eta},
\eea
where $\eta$ is a constant expected to be between 0.5 and 1.
The electron density can be parametrised as a $\beta$ model:
\bea
n_e(r)=n_0\left[1+\left(\frac{r}{r_c}\right)^2\right]^{-\frac{3}{2}\beta},
\eea
where $n_0$ is the central electron density, $r_c$ is the radius of the core itself, and $\beta$ is a constant. For a cool-core cluster such as the Perseus cluster, a double beta model can be used:
\begin{equation}\label{N}
n_e(r)=n_{0,1}\left[1+\left(\frac{r}{r_{c,1}}\right)^2\right]^{-\frac{3}{2}\beta_1}+n_{0,2}\left[1+\left(\frac{r}{r_{c,2}}\right)^2\right]^{-\frac{3}{2}\beta_2}.
\end{equation}
This paper extends previous studies on ALP constraints from the central AGN of the Perseus cluster to the case of massive ALPs. We first briefly review the dynamics of ALP-photon conversion.

\section{Axion-Photon Dynamics}
We align axes so that propagation is in the $z$ direction.
Following the derivation in \cite{RS}, the dynamics of the propagation of an axion-photon state $\ket{\Psi(E)}$ of energy $E$ is given by:
\begin{equation}\label{dyn}
\left[\omega+
\begin{pmatrix}
\Delta_\gamma & \Delta_F & \Delta_{g_{a\gamma\gamma},x} \\
\Delta_F & \Delta_\gamma & \Delta_{g_{a\gamma\gamma},y} \\
\Delta_{g_{a\gamma\gamma},x} & \Delta_{g_{a\gamma\gamma},y} & \Delta_a
\end{pmatrix}\right]
\begin{pmatrix}
\alpha_x \\
\alpha_y \\
\beta
\end{pmatrix}
=i\frac{\partial}{\partial z}
\begin{pmatrix}
\alpha_x \\
\alpha_y \\
\beta
\end{pmatrix},
\end{equation}
where $\ket{\Psi(E)}$ is a superposition state of $\ket{\gamma_x(E)},\ket{\gamma_y(E)}$ (the x- and y-polarized photon eigenstates, respectively), and $\ket{a(E)}$ (the axion eigenstate), as given by:
\bea\label{state}
\ket{\Psi(E)}=\alpha_x\ket{\gamma_x(E)}+\alpha_y\ket{\gamma_y(E)}+\beta\ket{a(E)}.
\eea
Here $\Delta_\gamma=-\frac{\omega^2}{2\omega_p}$, $\Delta_a=-\frac{m_a^2}{2\omega_p}$, $\Delta_{g_{a\gamma\gamma},x}=\frac{1}{2}B_xg_{a\gamma\gamma}$, $\Delta_{g_{a\gamma\gamma},y}=\frac{1}{2}B_yg_{a\gamma\gamma}$, 
and
\bea\label{plas}
\omega_p=\sqrt{\frac{4\pi\alpha n_e}{m_e}}
\eea
is the plasma frequency, for which $\alpha$ is the fine structure constant and $m_e$ is the electron mass.
$\Delta_F$ refers to Faraday rotation, which is negligible for X-ray energies and we set it to zero. In natural units, $\omega=E$. 

This equation can be used to evolve an initially pure photon state through many magnetic field domains and evaluate the conversion amplitude into an axion. During this evolution, the $\omega$ term on the left hand side of Eq. \ref{dyn} only contributes an overall phase into the final state, and thus may be ignored.\par

In the case of a single magnetic field domain of length $L$, the conversion probability $P(\ket{\gamma_x}\rightarrow\ket{a})$ simplifies to an illustrative analytic expression, 
\bea\label{conv}
P(\ket{\gamma_x}\rightarrow\ket{a})=4\vartheta^2\sin^2(\Delta_{osc}L/2),
\eea
where $\vartheta\approx\frac{1}{2}\tan(2\vartheta)=\frac{\Delta_{g_{a\gamma\gamma},x}}{\Delta_\gamma-\Delta_a}$ and $\Delta_{osc}=\Delta_\gamma-\Delta_a$.

\section{Source}
The photon source considered in this paper is the AGN of the central galaxy of the Perseus cluster NGC1275. Constraints in the case of massless axions were considered previously in \cite{ConGroup}, and this paper aims at extending bounds to the case of massive axions.\par
This source is particularly attractive from the perspective of constraining axions. It is extremely bright (one of the brightest extra-galactic sources in the X-ray sky). The brightness is useful because it results in a large number of photon counts, that gives statistical power in constraining any ALP-induced deviations away from an astrophysical power law. It is also located at the centre of a large, massive cool-core galaxy cluster - the Perseus cluster. This is advantageous as, on physical grounds, one expects the magnetic field environment of Perseus to be favourable for ALP-photon conversion. As a cool-core cluster, it has a high central magnetic field, and as a massive cluster one also expects a stronger magnetic field than for weaker, less developed clusters.

 The radial electron density profile $n_e(r)$ can be described by a double beta model of the form of Eq. \ref{N}, where
\begin{center}
\begin{eqnarray}
n_{0,1} & = & 3.9\times10^{-2}\SI{}{cm^{-3}}, \nn \\
n_{0,2} & = & 4.05\times10^{-3}\SI{}{cm^{-3}}, \nn \\
r_{c,1} & = & \SI{80}{kpc}, \nn \\
r_{c,2} & = & \SI{280}{kpc}, \nn \\
\end{eqnarray} 
\end{center}
and $\beta_1=1.2$, $\beta_2=0.58$ (\cite{Churazov2003}). 

Knowing $n_e(r)$, we may then infer an overall radial profile for the magnitude of the magnetic field $\lvert\vec{B}(r)\rvert$ using Eq. \ref{B}. We take an intermediary value of  $\eta=0.7$ and use $\lvert\vec{B}_0\rvert=\SI{25}{\mu G}$. This value is based on the central magnetic field estimated in \cite{Taylor2006} - for a different central value, the constraints on $g_{a\gamma\gamma}$ scale linearly with $B$ (larger $B$ allows weaker couplings to be excluded).\par

The electron density profile $n_e(r)$ also sets the radial profile of the plasma frequency $\omega_p(r)$ using Eq. \ref{plas}, and we plot this in Figure 1 (in practice, there may be localised fluctuations in the electron density, but here we treat this as smooth). We see that the plasma frequency declines from around $\omega_p\sim 10^{-11}\SI{}{eV}$ near the centre to $\omega_p \sim 10^{-12}\SI{}{eV}$ near the outskirts. 

This paper is concerned with axion masses in this range. Axion masses significantly lower than $10^{-12} {\rm eV}$ can be treated as effectively massless, whereas - as we shall see - masses significantly larger than $10^{-11} {\rm eV}$ are also uninteresting here, as then conversion is highly suppressed and no competitive bounds can be extracted.
\begin{figure}
\centering
\includegraphics[width=.45\textwidth]{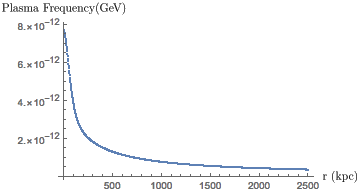}
\caption{Radial profile of plasma frequency $\omega_p$ for NGC1275}
\end{figure}

\section{Methodology}

The methodology used in this paper directly extends that described in \cite{ConGroup}, \cite{Marsh:2017yvc}, \cite{ConGroup2}, first developed in \cite{Wouters:2013hua} for their study of the central AGN of the Hydra A cluster. It involves taking
the spectrum of an AGN, fitting it to an absorbed power-law, and constraining the allowed couplings of axion-like particles by bounding any deviations from an astrophysical power-law.

\subsection{The Observed Data}
We use \textit{Chandra} observations of NGC1275. \textit{Chandra} is ideal for this purpose as its excellent angular resolution allows the AGN to be discriminated from the bright contaminating background of the Perseus cluster itself.
Although there are a total of 1.5Ms of \textit{Chandra} observations of NGC1275, we restrict to a subset of four with a total exposure of 200ks as detailed in table 1. There is no contamination from flares and we use all of the available time. 

For these four observations, the AGN is located around eight arcminutes away from the optical axis. For the remaining 1.3Ms, the AGN is on-axis for 1Ms and around three arcminutes off-axis for the remaining 300ks. The advantage of being so far off-axis is that the image is substantially spread out, greatly reducing the effects of pileup. In contrast, the on-axis observations are highly piled up, making them much less suited for the purpose of extracting a relatively clean spectrum of the AGN.

No observation can be entirely free of pileup. The effect of pileup is to worsen the quality of a fit, by redistributing photons to the wrong energies, with a distribution that is incompatible with either the spectrum of the source or the energy-dependent effective area of the telescope. 
Our bounds will be attained by excluding ALPs couplings that, when using data simulated using \texttt{fakepha}, give a fit clearly worse than the fit to the real data. As simulated data is cleaner than the real data, this process is conservative with respect to any residual pileup in the real data. Furthermore, it was shown in \cite{ConGroup} using MARX simulations for these off-axis observations that the uncertainties on bounds due to pile-up are much smaller than those due to the magnetic field uncertainty. For these reason we do not consider further any contamination due to the small residual effects of pileup.

The dataset was analysed using the analysis software CIAO 4.9 (together with  CALDB 4.7.4) and Sherpa (\cite{CIAO, sherpa}). After standard data reprocessing was applied, the spectrum was extracted from an ellipse of radii 7.636 and 5.240 arcseconds surrounding the AGN. The background was taken from a region surrounding and centered on the AGN consisting of a circular annulus which excluded the region occupied by the AGN itself.
\begin{table}
    \centering
    \begin{tabular}{ c|c|c|c }
    Obs ID & Exposure (ks) & Year & Instrument\\
    \hline\hline
    11713 & 112.24 & 2009 & ACIS-I   \\
    12025 &  17.93 & 2009 & ACIS-I   \\
    12033 &  18.89 & 2009 & ACIS-I   \\
    12036 &  47.92 & 2009 & ACIS-I   \\
    \end{tabular}
    \caption{The four \textit{Chandra} ACIS-I observations used. The counts/events from all four observations were later combined by the \textit{SHERPA} software, and then fit to a power law.}
\end{table}

For the data analysis, the extracted spectrum was restricted to the range $\SI{0.7}{keV}<E<\SI{5}{keV}$. The background was subtracted to reduce the number of counts present in the spectrum from the cluster thermal emission. We note that as the centre of the Perseus cluster is a rather complex and spatially inhomogeneous place, this will not entirely eliminate the contribution of thermal emission. However, given that before subtraction the thermal emission in the extraction region is not more than 10-15\% of the AGN emission (as shown in \cite{ConGroup}), subtraction does reduce it so that it is no more than a few per cent of the total emission. \par

The counts were then binned with 200 counts per bin, and the spectrum was fitted with an absorbed power-law (\texttt{xswabs*powlaw1d}),
\begin{eqnarray}
\label{powlaw}
P(E) & = & A E^{-\gamma}\times e^{-n_H\sigma(E)}.
\end{eqnarray}
Here $A$ is a normalisation factor, $\gamma$ is the power-law index, and $n_H$ is the hydrogen column density. The AGN is unobscured and, while one cannot exclude a small local contribution to the absorption, the dominant contribution comes from the Milky Way (which is large as Perseus is at low galactic latitude and close to the Milky Way disk). 

The fitting statistic used was the \texttt{chi2datavar} $\chi^2$ statistic of the \textit{SHERPA} fitting package provided with CIAO. The resulting fit is displayed in figure 2, and the fit parameters and reduced $\chi^2$ value are shown in table 2.
\begin{table}
    \centering
    \begin{tabular}{ c|c }
    Parameter & Value\\
    \hline\hline
    $\gamma$ & $1.82 \pm 0.01$   \\
    $n_H$ & $0.22 \pm 0.01$  \\
    reduced $\chi^2$ & $1.46$ \\
    \end{tabular}
    \caption{The power-law parameters (\ref{powlaw}) resulting from the fit, as well as the overall reduced $\chi^2$ value of the fit.}
\end{table}
\begin{figure}
\centering
\includegraphics[width=.45\textwidth]{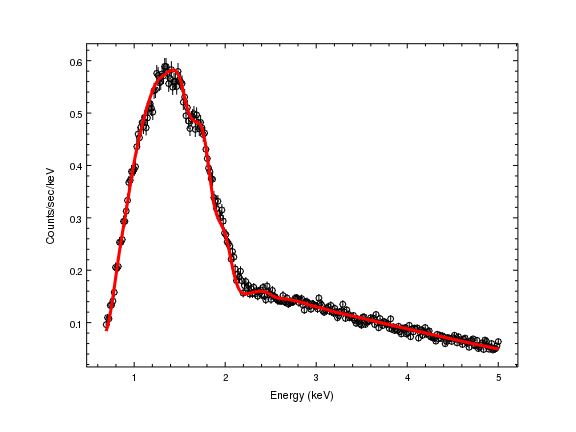}
\caption{Spectrum obtained for combined \textit{Chandra} ACIS-I observations of NGC1275, with 200 counts per bin, fitted to absorbed power law multiplied with resolution function using ``\texttt{chi2datavar}" statistics in \textit{SHERPA}.}
\end{figure}

As with previous work on this topic (\cite{Wouters:2013hua, ConGroup, ConGroup2, Marsh:2017yvc}), the ability to constrain ALPs arises from the fact that the absorbed power law is a reasonably good fit to the data (as can be seen in figure 2). ALP-photon couplings that would give deviations significantly larger than are actually observed in the data are therefore excluded. We now describe how we determine the expected level of modulations arising from ALPs of a specified mass and coupling.

\subsection{Propagation of ALP and Photon States}

The propagation of photons from NGC1275 to us, and their conversion into ALPs, is determined by the dynamics described in section 2. This depends on the precise form of the magnetic field within the Perseus cluster, as it determines the mixing between photon eigenstates and the axion state. 
However, while it may be possible to characterise statistically the overall strength of the magnetic field together with a reasonable estimate of its radial falloff, its exact configuration is unknown. The magnetic field arises from the turbulent multi-scale dynamics of the ICM, and so a precise magnetic field configuration along any line of sight is impossible to measure.\par

Thus, in order to simulate the propagation and time-evolution of a photon-axion state starting from the AGN source, we used (discrete) magnetic field configurations randomly generated in the following way. For each simulated magnetic field configuration, the total propagation length is split into 300 discrete domains, with lengths drawn randomly from the following distribution:
\begin{equation}
f(z)=\begin{cases}
0 & \text{if } z>10 \text{ or } z<3.5\\
N_0 z^{-2.2}& \text{if } 3.5\leq z\leq 10,
\end{cases}
\end{equation}
where $N_0$ is the appropriate normalisation constant and $z$ is the domain length given in units of kiloparsecs. 

Within each domain, the field is generated as a uniform magnetic field with magnitude set by Eq. \ref{B}, with $r$ evaluated as the distance from the near end of the domain to the center of the cluster. In any one domain, the direction $\theta$ of this magnetic field in the x-y plane is drawn from a flat distribution for $\theta\in[0,2\pi)$.\par 

For each value of $m_a$ and $g_{a\gamma\gamma}$, 100 such simulated magnetic fields were randomly generated. As per the notation of Eq. \ref{dyn} and \ref{state}, the x- and y-polarised states $\ket{\gamma_x}=$
$\begin{pmatrix}
1\\
0\\
0\\
\end{pmatrix}$
and 
$\ket{\gamma_y}=$
$\begin{pmatrix}
0\\
1\\
0\\
\end{pmatrix}$
were propagated using Eq. \ref{dyn} through each magnetic field realisation. This propagation was evaluated for 1000 different photon energies ranging from $0.01$ to $\SI{10}{keV}$. For each initial state, the photon survival probability was calculated from the final state
$\begin{pmatrix}
\alpha_x \\
\alpha_y \\
\beta
\end{pmatrix}$
as 
\bea
P(\ket{\gamma}\rightarrow\ket{\gamma})=\alpha_x^2+\alpha_y^2.
\eea
For a given energy $E$ and magnetic field configuration,
the overall survival rate is found by averaging the survival probabilities for both polarization states.\par

In this way, we determined the distribution of survival rates $P(E,\vec{B}(z),m_a,g_{a\gamma\gamma})$ for each of the 100 randomly generated magnetic fields $\vec{B}(z)$. We illustrate these survival rates in the figures below. The plots in Figure 3 were all generated using the same simulated magnetic field. From these, we can see clearly the way that larger masses increase the energy threshold necessary for there to be significant photon-ALP conversion.\par
\begin{figure}
\centering
    \includegraphics[width=.45\textwidth]{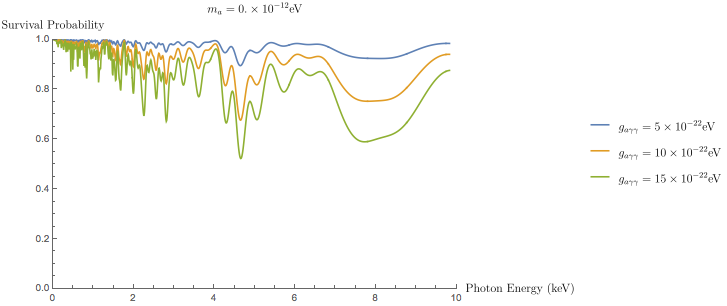}
    \includegraphics[width=.45\textwidth]{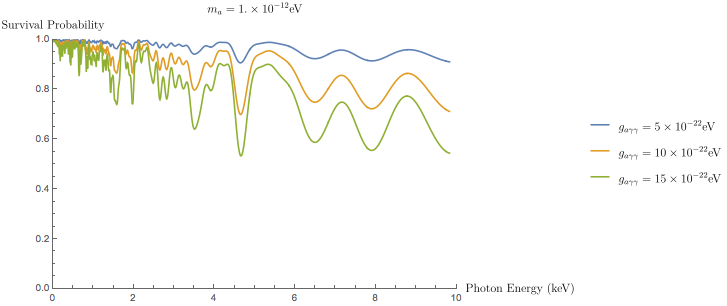}
    \includegraphics[width=.45\textwidth]{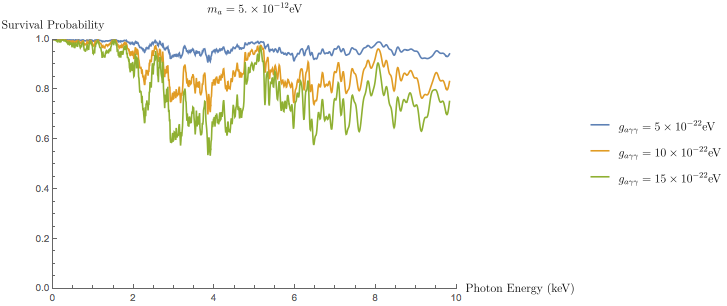}
    \includegraphics[width=.45\textwidth]{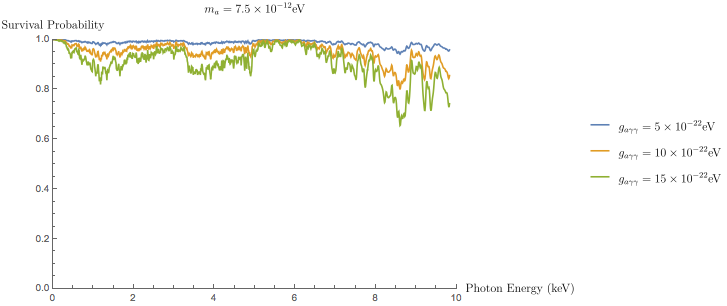}
    \includegraphics[width=.45\textwidth]{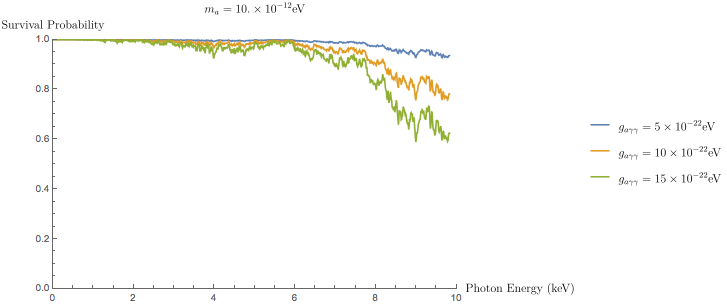}
    \caption{Photon survival probability vs. photon energy (given in units of keV) calculated for various values of  $g_{a\gamma\gamma}$ for various values of $m_a$ (both given in units of GeV) propagated through the same randomly-generated magnetic profile.}
\end{figure}
For the purposes of our project, this procedure was conducted for $\SI{0}\leq m_a\leq15\times10^{-12} \, \SI{}{eV}$ at increments of $0.5\times10^{-12} \, \SI{}{eV}$ and $\SI{0} \leq g_{a\gamma\gamma}\leq 1\times10^{-11} \, \SI{}{GeV}^{-1}$ at increments of $1\times10^{-13} \, \SI{}{GeV}^{-1}$.

\subsection{The Simulated Data}
In order to determine the values of $m_a$ and $g_{a\gamma\gamma}$ for which the resulting photon-axion conversions (and the spectral modulations they induce) are too great to be consistent with the observed data, we carried out the following procedure.

For every fixed value of $m_a$ and $g_{a\gamma\gamma}$, the survival probability distribution for each of the 100 randomly-generated magnetic fields was loaded into \textit{SHERPA} as a table model, and then multiplied by the absorbed power law described in section 4.1. This gives a `fake' model representing the arrival of spectrum of photons, in the case of an ALP of specified mass and coupling.

Using \textit{SHERPA}'s \texttt{fakepha} command, this model was then used to generate simulated `fake' spectra, corresponding to those that would have been observed by \textit{Chandra} on propagation through the magnetic field model and in the case that ALPs actually existed with the specified masses and couplings $m_a$ and $g_{a\gamma\gamma}$.

In the case of large coupling, this simulated data should be a bad fit. To measure this, we re-fitted it to a pure power law, obtaining the reduced $\chi^2$ value of the fit. 
We then compared this reduced $\chi^2$ to that obtained when fitting the actual $\textit{Chandra}$ data. For each pair of $m_a$ and $g_{a\gamma\gamma}$, we record the percentage of simulated spectra with worse fits (i.e. a higher reduced $\chi^2$ value) than that obtained in the fit to the actual data. The higher this percentage, the less acceptable the particular ($m_a$, $g_{a\gamma\gamma}$) pair is. The results are displayed in figure 4, for exclusions of $67\%$, $95\%$, and $99\%$ (a $95\%$ exclusion means that $95\%$ of simulated data samples gave worse fits than the actual data).

\section{Analysis and Conclusions}

The results of Figure 4 provide upper bounds on the ALP-photon coupling $g_{a\gamma\gamma}$ for every axion mass $m_a$. This plot contains three main regions of interest.

The first is that corresponding to the large mass limit, roughly $m_a \gtrsim 9 \times10^{-12}\SI{}{eV}$. As illustrated in Figure 3, at high masses photon-axion rapidly becomes highly suppressed, and conversion probabilities fall off as $g_{a \gamma \gamma}^2 m_a^{-4}$. This accounts for the steep rise in the allowed region in this region -
when survival probabilities are close to unity across all energies, there are no modulations and so the simulated data and observed data are indistinguishable. In this region, the exclusion limits on $g_{a\gamma\gamma}$ rapidly become weak and uncompetitive with other constraints.

The second nontrivial region of interest is the small-mass region corresponding to $0\lesssim m_a\lesssim6\times10^{-12}\SI{}{eV}$. Here, three roughly uniform bounds may be set on $g_{a\gamma\gamma}$ corresponding to the three exclusion levels we are considering ($33\%$, $5\%$, and $1\%$). Here, the $67\%$ (respectively $95\%$ and $99\%$) exclusion limits are $g_{a\gamma\gamma} \lesssim 1.20 (1.55, 1.75) \times 10^{-12}\SI{}{GeV^{-1}}$. This is consistent with the results found for massless ALPs in the analysis of \cite{ConGroup}, which this paper generalises.

The third region of interest is that contained within $6 \times 10^{-12}\SI{}{eV}\lesssim m_a \lesssim 9.0 \times 10^{-12}\SI{}{eV}$. As discussed in section 2, in this region,
as the photons pass through the Perseus cluster they go through regions where the axion mass $m_a$ is identical to the plasma frequency $\omega_p$.  This manifests itself within the photon-axion conversion through the probabilities displayed in figure 3 for $m_a=7.5\times10^{-12}\SI{}{eV}$, in which significant conversion occurs even at lower photon energies $E$, unlike the probabilities calculated for the other masses displayed. Thus, while the $33\%$ exclusion level gives a bound of roughly $g_{a\gamma\gamma}\lesssim1.2\times10^{-12}\SI{}{GeV^{-1}}$, equal to that established for region 2, the bounds for the $5\%$, and $1\%$ exclusion levels are higher than those established for region 2, appearing as a brief spike. This peaks at $m_a\approx7.5\times10^{-12}\SI{}{eV}$, for which the $5\%$ exclusion bound is $g_{a\gamma\gamma}\lesssim2.5\times10^{-12}\SI{}{GeV^{-1}}$, with an additional ``island" at roughly $2.75\times10^{-12}\SI{}{GeV^{-1}}\lesssim g_{a\gamma\gamma}\lesssim3.0\times10^{-12}\SI{}{GeV^{-1}}$, and for which the $1\%$ exclusion bound is $g_{a\gamma\gamma}\lesssim3.75\times10^{-12}\SI{}{GeV^{-1}}$.

In summary, we have extended the exclusion limits on massless or ultralight ALPs obtained in \cite{ConGroup} to the case of massive ALPs. For cases of an ALP mass $5 \ti 10^{-12} \, {\rm eV} \lesssim m_a \lesssim 10 \ti 10^{-12} \, {\rm eV}$, this offers new and competitive constraints. Looking to the future, it is clear that in the regime of $m_a \lesssim 10^{-11} {\rm eV}$ X-ray astronomy offers the most competitive methods to constrain (or discover) light ALPs.

\begin{figure}
\centering
\includegraphics[width=.45\textwidth]{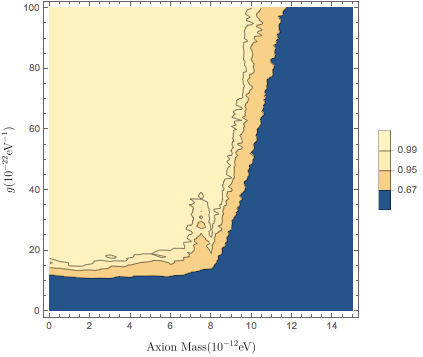}
\caption{Contour plot plotting percentage of better fits against $m_a$ and $g_{a\gamma\gamma}$}
\end{figure}

\begin{figure}
\centering
    \includegraphics[width=.45\textwidth]{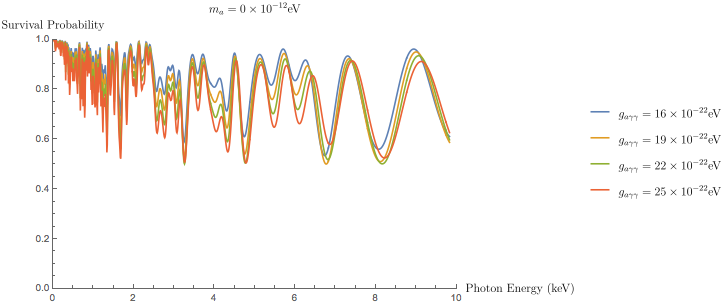}
    \includegraphics[width=.45\textwidth]{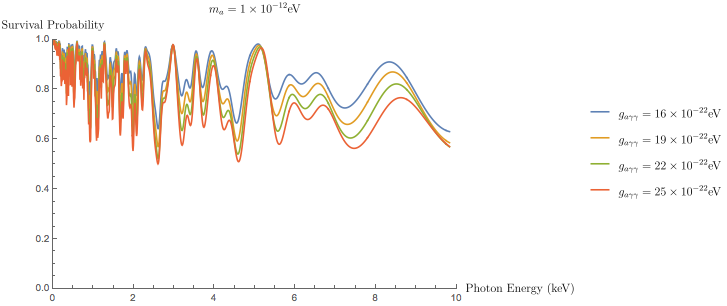}
    \includegraphics[width=.45\textwidth]{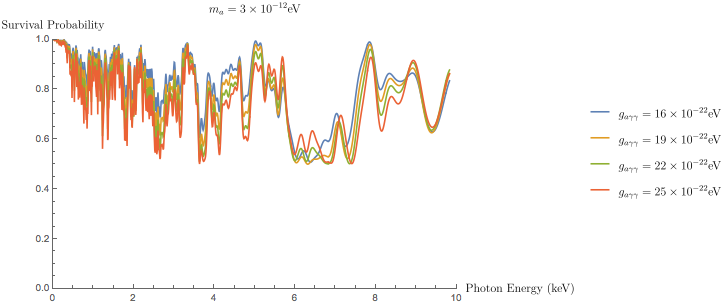}
    \includegraphics[width=.45\textwidth]{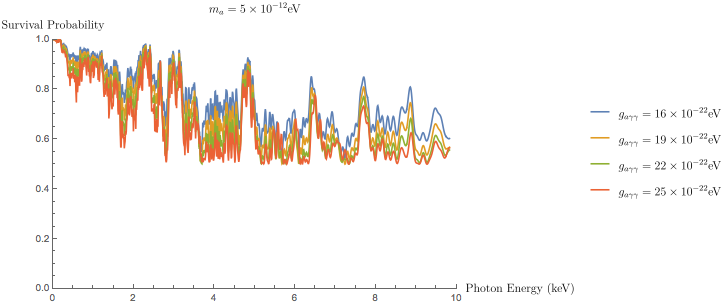}
    \caption{Photon survival probability vs. photon energy (given in units of keV) calculated for various values of  $g_{a\gamma\gamma}$ for various values of $m_a$ (both given in units of GeV) within the $5\%$ exclusion zone, propagated through the same randomly-generated magnetic profile.}
\end{figure}

\section{Acknowledgements}
LC thanks Princeton University and its International Internship Program (IIP) for setting up this summer research studentship and for funding, as well as Luisa Duarte-Silva and Michelle Bosher for handling 
logistical matters at Princeton and Oxford respectively. He also thanks Steve Gubser for requesting that such a program be made possible. He also thanks Francesca Day, Nick Jennings and Sven Krippendorf for assistance and conversations. JC was supported by a European Research Council Starting Grant `Supersymmetry Breaking in String Theory' (307605).




\bibliographystyle{mnras}
\bibliography{main}



\appendix

\bsp	
\label{lastpage}
\end{document}